# Formation of stationary electronic states in finite homogeneous molecular chains

V.D. Lakhno,[*] A.N. Korshunova[†]

*Institute of Mathematical Problems of Biology*
*Russian Academy of Sciences, Pushchino, Moscow Region, 142290, Russia*

**Abstract**

Evolution of an arbitrary initial distribution of a quantum-mechanical particle in a uniform molecular chain is simulated by a system of coupled quantum-classical dynamical equations with dissipation. Stability of a uniform distribution of the particle over the chain is studied. An asymptotical expression is obtained for the time in which a localized state is formed. The validity of the expression is checked by direct computational experiments. It is shown that the time of soliton and multisoliton type states formation depends strongly on the initial phase of the particle's wave function. It is shown that in multisoliton states objects with a fractional electron charge which can be observed experimentally are realized. The results obtained are applied to synthetic uniform polynucleotide molecular chains.

[*]*E-mail: lak@impb.psn.ru*
[†]*E-mail: alya@impb.psn.ru*

## 1. INTRODUCTION

We witness an ever increasing interest of physicists, chemists and biologists in the problem of conducting properties of molecular chains which are considered a promising matter to be used in nanoelectronics [1, 2, 3, 4, 5]. In this case, in an effort to interpret theoretically the results of measurements of the conductivity or the charge transfer rate one inevitably faces the problem of the nature of charge carriers in such chains. The main candidates for the role of charge carriers in such quasionedimensional system as molecular chains are solitons and polarons [6, 7], [8, 9, 10], [11, 12, 13, 14, 15]. Here, however, of importance is the time in which such states are formed in such chain. Thus, for example, if the time in which a soliton is formed in a short chain is longer than the time of transfer, the transfer process will have a complicated nonstationary character and will immensely depend on the initial conditions of the charge density distribution on the contacts and neighboring molecules. Accordingly, one might expect a great dispersion of the experimental results and their non-reproducibility.

In this work we simulate the dynamics of the formation by quantum particle (electron or hole) of a standing soliton states in a chain consisting of $N$-sites representing harmonical oscillators. Though being simple, the model provides the basis for the description of solid-state crystals in which deviations of the lattice atoms from their equilibrium positions are described in terms of harmonical oscillators.

In this paper we study possible stationary quantum dissipative structures which are looking like multisoliton states , and the time required for the formation of a soliton states in a molecular chain, since this parameter is of crucial importance in elucidating the nature of charge carriers in the chain. Earlier [16, 17] we considered this problem for a model molecular chain. Here we partly reproduce the results of [16, 17] corresponding to the case of a completely relaxed scenario of a soliton formation. However, if the condition of a complete relaxation is not fulfilled, formation of a soliton proceeds by quite a different mechanism. Such a situation can take place in polynucleotide chains and is considered here in detail.

In recent years, localization of energy in nonlinear lattices has been the subject of intensive theoretical and experimental investigations [18, 19, 20, 21, 22]. Many of them have dealt with the dynamics of Discrete Nonlinear Schrödinger (DNLS) equation. That is why we believe that the results obtained here may be of interest for a wide range of problems such as local denaturation of the DNA double strand [23], self-trapping of vibrational energy in proteins [24], propagation of discrete self-trapped beams in arrays of coupled optical waveguides [25, 26], nonlinear excitations in hydrogen-bounded crystals [27] etc.

Fundamentally, the problems discussed in this work are likely to fall within the domain of vigorously progressing nonlinear science, such as dissipative solitons [28]. As distinct from [28], where classical models are used, our case of an electron (hole) interacting with a classical chain with dissipation is an example of a quantum-classical dissipative system.

The paper is arranged as follows.
In section 2 we introduce a Hamiltonian of a charge in molecular chain where the motion of a charge is considered quantum-mechanically and that of molecular chain - classically, with regard for a dissipation in the classical subsystem.
In section 3 quantum-classical dissipative systems are considered in general. It is shown that if a system exhibits dissipation, the quantum subsystem evolves to its ground state.
In section 4 we deal with a rigid uniform chain in which any deformations are lacking. Here is applicable a concept of a band structure of a uniform chain in which eigen wave functions

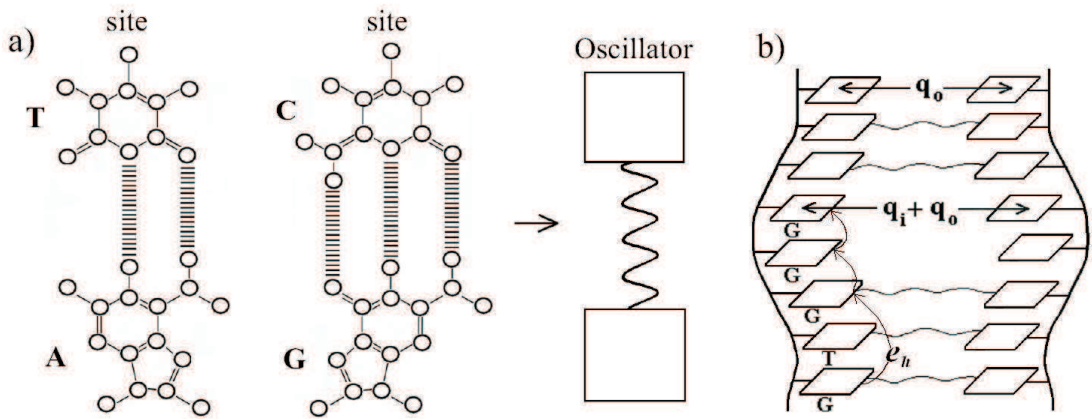

**Fig. 1.** Schematic representation of Watson-Crick pairs as harmonical oscillators - a) and charge transfer along a nucleotide chain - b).

describe the states of a charge delocalized over the chain.

In section 5 we show that in a uniform deformable chain, with excess charge placed in it, localized standing soliton-type states, arising without overcoming a potential barrier, have a lower energy than delocalized ones.

In section 6 we obtain an asymptotic estimation of the time in which a localized state is formed from an initially delocalized state of a charged particle in a uniform molecular chain with dissipation.

In section 7 in the course of direct computational experiments we model the process of the formation of a localized state from delocalized states of various types. It is shown that the time in which a localized state becomes steady depends greatly on the initial phase of the wave function of a quantum particle. The validity of the asymptotical estimates obtained in the previous section is demonstrated.

In section 8 we consider formation of excited states of a multisoliton-type particle. The considered case represents an example of self-organization from "unlimited" waves steady localized particle-like structures with a fractional charge.

In section 9 the results obtained for molecular chains in the previous sections are applied to polynucleotide chains.

Section 10 is devoted to a discussion of the results obtained.

In the Appendix we study the stability of a uniform charge distribution in a molecular chain.

## 2. HAMILTONIAN AND DYNAMICAL EQUATIONS

In the model discussed below, the molecular is considered as a chain of $N$ sites. Each site represents a pair of atoms, molecules or molecular complex which is treated as a harmonical oscillator. There is placed an excess charge (electron or hole) capable of moving over the chain. As an example we refer to a polynucleotide chain in which a nucleotide pair is considered as a site and a radical cation or a hole formed in it is treated as an excess charge (Figure 1). To model the dynamics of a particle in a system of $N$-sites representing independent oscillators we will proceed from Hamiltonian by Holstein ($\hat{H}$) who was the first to consider a chain in which each site represents a biatomic molecule [6, 7, 9, 29]:

$$\hat{H} = \hat{H}_Q + H_{cl}, \quad \hat{H}_Q = \sum_{n,m}^N v_{nm}|n\rangle\langle m|, \quad v_{nn} = v_n = \alpha_n^0 + \alpha_n' q_n,$$
$$H_{cl} = T_k + U_p, \quad T_k = \sum_n^N M_n \dot{q}_n^2/2, \quad U_p = \sum_n^N k_n q_n^2/2, \quad (1)$$

where $\hat{H}$ - is the Hamiltonian of a particle interacted with chain deformation, $v_n$ - is the energy of the particle on the $n$-th site with the wave function $|n\rangle$, $v_{nm}(n \neq m)$ -are matrix elements of the particle transition from the $n$-th site to the $m$-th site, $T_k$ is the kinetic energy of the oscillator sites, $M_n$ is the effective mass of an oscillator, $q_n$ is a deviation of a site from its equilibrium position, $U_p$ is the potential energy of the oscillators, $k_n$ is an elastic constant, $v_{nn}$ - are the energies of particle on $n$-th site, $\alpha_n'$ - are particle-site displacement coupling constants.

We will seek a solution of the wave equation corresponding to Hamiltonian (1) in the form:

$$|\Psi(t)\rangle = \sum_{n=1}^N b_n(t)|n\rangle. \quad (2)$$

Motion equations for Hamiltonian $\hat{H}$ in the nearest neighbor approximation, yield the following system of differential equations:

$$i\hbar \dot{b}_n = \alpha_n^0 b_n + \alpha_n' q_n b_n + v_{n,n-1} b_{n-1} + v_{n,n+1} b_{n+1}, \quad (3)$$

$$M_n \ddot{q}_n = -\gamma_n \dot{q}_n - k_n q_n - \alpha_n' |b_n|^2. \quad (4)$$

Equations (3) are Schrödinger equations for the probability amplitudes $b_n$, describing evolution of a particle in a deformable chain, where $\hbar = h/2\pi$, $h$ - is a Plank constant, while equations (4) are classical motion equations describing the site dynamics with regard for dissipation, where $\gamma_n$ - is a friction coefficient of the $n$-th oscillator.

In equations (3),(4) we pass on to dimensionless variables with the use of the following relations:

$$\eta_n = \tau \alpha_n^0/\hbar, \quad \eta_{nm} = \tau v_{nm}/\hbar, \quad \tilde{\omega}_n^2 = \tau^2 k_n/M_n, \quad \omega_n' = \tau \gamma_n/M_n, \quad (5)$$
$$q_n = \beta_n u_n, \quad \varkappa_n \tilde{\omega}_n^2 = \tau^3 (\alpha_n')^2/\hbar M_n, \quad \beta_n = \tau^2 \alpha_n'/M_n, \quad t = \tau \tilde{t},$$

where $\tau$ is an arbitrary time scale relating the time $t$ and a dimensionless variable $\tilde{t}$.

In terms of dimensionless variables (5), equations (3),(4) are written as:

$$i\frac{db_n}{d\tilde{t}} = \eta_n b_n + \eta_{n,n+1} b_{n+1} + \eta_{n,n-1} b_{n-1} + \varkappa_n \tilde{\omega}_n^2 u_n b_n, \quad (6)$$

$$\frac{d^2 u_n}{d\tilde{t}^2} = -\omega_n' \frac{du_n}{d\tilde{t}} - \tilde{\omega}_n^2 u_n - |b_n|^2. \quad (7)$$

In the nearest neighbor approximation, the total energy $E = \langle\Psi|\hat{H}|\Psi\rangle$, corresponding to Hamiltonian (1), has the form:

$$\tilde{E} = \tilde{H}_Q + \tilde{H}_{cl},$$
$$\tilde{H}_Q = \sum_n \eta_n |b_n|^2 + \sum_n \varkappa_n \tilde{\omega}_n^2 u_n |b_n|^2 + \sum_n (\eta_{n,n+1} b_n^* b_{n+1} + \eta_{n,n-1} b_n^* b_{n-1}) \quad (8)$$
$$\tilde{H}_{cl} = \frac{1}{2}\sum_n \varkappa_n \tilde{\omega}_n^2 \dot{u}_n^2 + \frac{1}{2}\sum_n \varkappa_n \tilde{\omega}_n^4 u_n^2,$$

where $\widetilde{E} = E\tau/\hbar, \widetilde{H}_Q = H_Q\tau/\hbar, \widetilde{H}_{cl} = H_{cl}\tau/\hbar$ - dimensionless values of $E, H_Q, H_{cl}$.

Thus, we have introduced a simplest model to describe the dynamics of a quantum particle in a classical chain where the dissipation in the system is taken into account in an explicit form.

In the subsequent discussion of utility will be the following property of the system (6),(7): distributions of the probabilities $|b_n(t)|^2$ of a particle occurrence on the sites, obtained as a result of solution of (6),(7), do not depend on the sign of the matrix elements $\eta_{n,n-1}, \eta_{n,n+1}$ [30]. To prove this statement let us assume:

$$b_n = e^{in\pi}\tilde{b}_n. \tag{9}$$

Substitution of (9) into (6),(7) yields equations for the amplitudes $\tilde{b}_n$, which differ from (6),(7) only in the sign of $\eta_{n,n-1}$ and $\eta_{n,n+1}$. So, a change in the sign of the matrix elements alters only the phase of the wave function amplitude and leaves unaltered the quantities $|b_n(t)|^2$, and thus does not change the physical results obtained.

## 3. EVOLUTION OF QUANTUM-CLASSICAL DISSIPATIVE SYSTEMS

We can make some general statements about the quantum-classical system (3),(4) on which our further consideration will be based. In a purely quantum system described by the Hamiltonian:

$$\hat{H}_Q = \sum_{n,m}^{N} v_{nm}|n\rangle\langle m|, \tag{10}$$

an irreversible evolution does not take place.

The wave function (2) determines the energy of the quantum system $E_Q$:

$$E_Q = \langle\Psi|\hat{H}_Q|\Psi\rangle = \sum_{n,m}^{N} v_{nm}b_n^*(t)b_m(t). \tag{11}$$

which is independent of time: $E_Q(t) = E_Q(t_0)$.

For an irreversible evolution to arise, in addition to a quantum system, a classical system must be available and these systems must interact. With the availability of a classical system, the energy of a quantum system $E_Q$ will depend on the variables of the classical system $q_1 \ldots q_k \ldots$ as on the parameters: $E_Q(t) = E_Q(q_1(t), \ldots q_k(t), \ldots)$, which provides an interaction of the quantum and classical systems.

In the case of a classical system, the motion equations of the Hamiltonian will take the form:

$$\dot{q}_n = \frac{\partial H_{cl}}{\partial P_n}, \tag{12}$$

$$\dot{P}_n = -\frac{\partial H_{cl}}{\partial q_n} - \frac{\partial E_Q}{\partial q_n} - \frac{\partial F}{\partial \dot{q}_n}, \tag{13}$$

where: $H_{cl} = H_{cl}(P_1, \ldots P_k, \ldots q_1, \ldots, q_k \ldots)$ - is the Hamiltonian of the classical system. These equations involve the dissipative function of the classical system F. A change in the energy of a quantum system with time is written as:

$$\dot{E}_Q = \sum_n \dot{q}_n \frac{\partial}{\partial q_n} E_Q(q_1, \ldots, q_k \ldots) + \frac{\partial E_Q}{\partial t}. \tag{14}$$

Accordingly, a change in the energy of the classical system $H_{cl}$ with time is:

$$\frac{dH_{cl}}{dt} = \frac{\partial H_{cl}}{\partial t} + \sum_n \frac{\partial H_{cl}}{\partial P_n}\dot{P}_n + \sum_n \frac{\partial H_{cl}}{\partial q_n}\dot{q}_n = \quad (15)$$

$$\frac{\partial H_{cl}}{\partial t} + \sum_n \dot{q}_n \left(\dot{P}_n + \frac{\partial H_{cl}}{\partial q_n}\right) = \quad (16)$$

$$\frac{\partial H_{cl}}{\partial t} - \sum_n \dot{q}_n \frac{\partial E_Q}{\partial q_n} - \sum_n \dot{q}_n \frac{\partial F}{\partial \dot{q}_n}, \quad (17)$$

namely:

$$\frac{dH_{cl}}{dt} = -\frac{dE_Q}{dt} + \frac{\partial E_Q}{\partial t} - \sum_n \dot{q}_n \frac{\partial F}{\partial \dot{q}_n} + \frac{\partial H_{cl}}{\partial t}. \quad (18)$$

Then a change in the total energy of the quantum and classical systems: $E = E_Q + H_{cl}$ will be:

$$\begin{aligned}\frac{dE}{dt} &= \sum_n \dot{q}_n\left(\dot{P}_n + \frac{\partial H_{cl}}{\partial q_n}\right) + \frac{\partial H_{cl}}{\partial t} + \sum_n \dot{q}_n \frac{\partial}{\partial q_n}E_Q + \frac{\partial E_Q}{\partial t} = \\ &= -\sum_n \dot{q}_n \frac{\partial F}{\partial \dot{q}_n} + \frac{\partial E_Q}{\partial t} + \frac{\partial H_{cl}}{\partial t}.\end{aligned} \quad (19)$$

In the case of a purely mechanical system, the dissipative function is equal to:

$$F = \frac{1}{2}\sum_{n,m} \gamma_{nm} \dot{q}_n \dot{q}_m. \quad (20)$$

Equation (19) suggests that the tendency of the evolution of a quantum-mechanical system is determined by the classical system. As mentioned earlier, this follows from the fact that a quantum system not interacting with a classical one is time reversible. Irreversibility arises when a quantum system starts interacting with the classical system under consideration. In this case the classical system determines the tendency of the evolution while the interaction between the quantum and classical systems governs the rate of this evolution.

## 4. UNIFORM RIGID CHAIN: $\alpha' = 0$

In describing how a localized state of a particle is formed in a uniform chain with the parameters: $\alpha'_n = \alpha'$, $\nu_{n,n\pm 1} = \nu$, $M_n = M$, $k_n = k$, $\gamma_n = \gamma$, $\alpha^0_n = \alpha^0 = 0$, we will proceed from the fact that initially a particle is delocalized and the characteristic size of this delocalized state is of the order of the chain length. This situation arises when a particle gets into an undeformed chain. Thus, for example, in a rigid chain with $\alpha' = 0$ equation (3) has the solution:

$$b_n(t) = \sum_m b_m(0)(-i)^{n-m} J_{n-m}(2\nu t/\hbar), \quad (21)$$

where $b_m(0)$ - is the value of the wave function amplitude at the moment $t = 0$, $J_n$ - is the Bessel function of the first kind. From (21) it follows that a particle, being localized at an initial moment $t = 0$ at the site with number $n = 0$, immediately acquires a nonzero probability to occur everywhere over the chain at a time (the probability decreases at large $n$ as $n^{-1}$). Under numerical solution of equation (3),(4) this manifests itself in the fact that a particle, being placed at an initial moment at one of the sites, in no time "spreads" over the whole length of the chain.

Notice, that the solution of stationary Schrödinger equation (3) for the Hamiltonian $\hat{H}$ determines the band of permitted energies $W_k$:

$$W_k = 2\nu \cos k, \ k = 2\pi l/N, \ l = 0, \pm 1, \pm 2, \ldots, \pm N/2, \quad (22)$$

which correspond to the wave functions:

$$b_{nk}(t) = e^{-iW_k t} \cdot e^{ikn}/\sqrt{N+1}. \quad (23)$$

According to (22), at $\nu < 0$ the lowest energy corresponds $k = 0$, at which $W_0 = 2\nu = -2|\nu|$, and at $\nu > 0$ it corresponds to $k = \pm\pi$, at which $W_{\pm\pi} = -2\nu$. So, in this case transformation (9) determines a correlation between the wave functions corresponding to the minimal and maximal energies of a particle, such that if $b_n$ corresponds to the minimal energy, then $\tilde{b}_n$ - to the maximal one, and vice versa.

## 5. UNIFORM DEFORMABLE CHAIN: $\alpha' \neq 0$

At $\alpha' \neq 0$ a particle, being delocalized, starts gradually deforming the chain and, after a lapse of rather a long time, it turns into a localized state. To find the time in which it becomes localized let us consider the case when in the course of localization the particle distribution density $|b_n(t)|^2$ changes slowly so that the chain has time to relax completely. This situation is determined by the condition:

$$|\nu| \ll \hbar\omega' \lesssim \hbar\omega, \quad (24)$$

at which equation (4) has the approximate solution:

$$q_n(t) = -\frac{\alpha'}{k}|b_n(t)|^2. \quad (25)$$

In this case equation (3) takes the form of the nonlinear Schrödinger equation:

$$i\hbar \dot{b}_n = \nu(b_{n-1} + b_{n+1}) - \frac{\alpha'^2}{k}|b_n|^2 b_n. \quad (26)$$

In the continuum approximation, when the wave function $b_n$ smoothly changes with changing $n$, equation (26) turns to:

$$i\hbar \dot{b}_n = \nu \frac{\partial^2 b_n}{\partial n^2} - \frac{\alpha'^2}{k}|b_n|^2 b_n. \quad (27)$$

Equation (27) can be obtained by variation of the functional $E_\nu(t)$:

$$E_\nu(t) = 2\nu - \nu \int \left|\frac{\partial b_n(t)}{\partial n}\right|^2 dn - \frac{\alpha'^2}{2k} \int |b_n(t)|^4 dn, \quad (28)$$

with respect to $b_n$. Notice that this functional has the meaning of the total energy.

The case of $\nu < 0$ corresponds to the functional $E_\nu(t)$ limited from below, the minimal value of which corresponds to a stable stationary state $b_n^0$ such that:

$$b_n^0 = \frac{\sqrt{2}}{4}\sqrt{\frac{\alpha'^2}{k\nu}} \cdot \exp i\varphi \cdot \text{ch}^{-1}\left(\frac{\alpha'^2}{4k\nu}(n-n_0)\right), \tag{29}$$

representing a standing soliton.

At $\nu > 0$ functional (28) is not limited from below and any initial state is unstable.

## 6. TIME REQUIRED FOR THE FORMATION OF A STANDING SOLITON STATE

To describe the transition of a particle from a delocalized state to the localized one of the form of (29) let us introduce the function:

$$b_n(t) = \xi^{1/2}(t) b_0(\xi(t)n), \tag{30}$$

which satisfies the normalizing condition:

$$\int |b_n(t)|^2 dn = 1, \tag{31}$$

and at $t = -\infty$, represents a delocalized state:

$$\xi(-\infty) = 0. \tag{32}$$

Substituting wave function (30) into functional (28) we get:

$$E(t) = \xi^2(t)A - \xi(t)B + 2\nu, \tag{33}$$

$$A = -\nu \int \left|\frac{\partial b_n(t)}{\partial n}\right|^2 dn, \; B = \frac{\alpha'^2}{2k}\int |b_n(t)|^4 dn,$$

where $A$ and $B$ do not depend on t.

Functional (28) reaches its minimal value (in the case of $\nu < 0$) when a standing soliton state is formed at a moment $t_s$. In this case:

$$\xi(t_s) = 1. \tag{34}$$

To find the time $t_s$ let us consider the process of a soliton formation in a dissipative medium determined by the dissipation function F:

$$F(t) = \frac{1}{2}\gamma \int \dot{q}^2(n,t) dn, \tag{35}$$

where $\gamma$ is a dissipation coefficient, and $q(n,t) = q_n(t)$ at each instant of time is determined by function (30) in accordance with relation (25):

$$q(n,t) = -\frac{\alpha'}{k}\xi(t)\left|b_0(\xi(t)n)\right|^2. \tag{36}$$

Substitution of (28), (35), (36) into the energy distribution equation which describes a relation between the energy assumed by a particle as a result of its localization and that lost due to dissipation:

$$\dot{E} = \int \dot{q}(n,t) \frac{\delta F}{\delta \dot{q}(n,t)} dn, \qquad (37)$$

yields the following equation for the parameter $\xi$:

$$\dot{\xi} = \alpha_s \xi (1-\xi), \ \alpha_s = ck/\gamma, \qquad (38)$$

where $c$ is a constant of the order of one. Equation (38) satisfying condition (32) and the condition $\xi(t_s) = 1$ has a solution only in the case of $t_s = \infty$:

$$\xi(t) = p \exp \alpha_s t / (1 + p \exp \alpha_s t), \qquad (39)$$

where $p$ is an arbitrary constant.

So, a soliton state is formed from an initial delocalized state in an infinite time. In a real system the parameter $\xi$ is never equal to zero at the initial moment. Since $\xi^{-1}$ has the meaning of the characteristic size of the state, the value of $\xi^{-1}$ is always limited by the size of the system, and in a nonideal chain — by the characteristic size of the interval on which the chain has some distortions. Suppose $\xi(0) \to 0$, which corresponds to the initial delocalized state. From (39) follows that $p = \xi(0)$, with an accuracy of the terms of the order of $\xi^2(0)$. So, from (39) follows that $\xi(t_s) = \xi(0) e^{\alpha_s t_s}/(1 + \xi(0) e^{\alpha_s t_s})$ and $\xi(t_s) \sim 1$ for $\xi(0) e^{\alpha_s t_s} \gtrsim 1$. Hence, the time of the soliton formation is assesses as $t_s \sim (1/\alpha_s) \ln(1/\xi(0))$, i.e.

$$t_s = \frac{\gamma}{k} \ln(1/\xi(0)). \qquad (40)$$

In terms of dimensionless variables (5) this relation is written as:

$$\tilde{t}_s \sim \frac{\omega'}{\omega^2} \ln(1/\xi(0)). \qquad (41)$$

## 7. NUMERICAL MODELING OF A SOLITON FORMATION AT VARIOUS INITIAL PARAMETER VALUES

Here we present the results of direct computational experiments on determining the dependence of the soliton formation time on the initial distribution of a particle and the values of the parameters $\omega^2$, $\omega'$. The calculations were carried put by standard numerical methods for solution of the systems of nonlinear differential equations, i.e. 4-th order Runge-Kutta method and Adams (explicit and implicit) methods.

Let us choose the sequence length to be $Ns = 51$, and values of $\varkappa_n = \varkappa = 2$, $\eta = 0.8$ ($\eta > 0$), where $\eta = \tau \nu/\hbar$, $\tau = 10^{-14} sec$. At these parameter values one soliton is formed at various $\omega^2$, $\omega'$ and various initial values of $x_n$, $y_n$ ($b_n = x_n + iy_n$). Let us write $T_{sol}$ for the time in which a soliton becomes steady. Figure 2 presents the graphs of the functions $|b_n|^2$, $x_n$, $y_n$, for a steady soliton at the above-indicated parameter values.

Let us see how the time of a soliton formation depends on the initial values of $x_n$, $y_n$ at fixed values of $\omega^2$, $\omega'$. Let us take $\omega^2 = \omega' = 1$, $du_n^0/dt = 0$, and $u_n^0$ related by (25) to the initial values of $|b_n(0)|^2$ such that $u_n^0 = |b_n(0)|^2/\omega_n^2$. Let us calculate the functions $|b_n|^2$, $x_n$, $y_n$, for an a fortiori steady soliton, that is for the time $T_{sol} \to \infty$, using the above-indicated parameter values. We will compare the graphs of the functions $|b_n(t)|^2$ and $|b_n(T_{sol})|^2$ in order to understand in what manner and how quickly the distribution of a particle over the chain will take the form of a soliton.

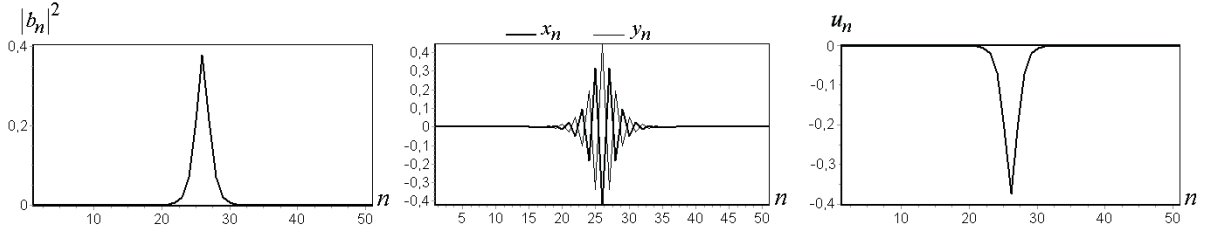

**Fig. 2.** Graphs of the functions $|b_n|^2$ ($b_n = x_n + iy_n$), $x_n$, $y_n$, $u_n$ for the time $t > T_{sol}$.

Let us introduce the function $\delta(t) = \sum_{n=1}^{Ns} (|b_n(t)|^2 - |b_n(T_{sol})|^2)^2$. The function $\delta(t)$ represents a mean-square deviation of $|b_n(t)|^2$ from the initial distribution function.

Let us consider various initial distributions.

### Uniform initial distribution

The problem of the stability of a uniform charge distribution in general case, given by equations (3),(4), is considered in the Appendix.

Let us consider the case when the initial values of the amplitudes of the probabilities of a charge occurrence on some or other site are similar for all the sites: $|b_n(0)| = 1/\sqrt{Ns}$, $\sum_{n=1}^{Ns} |b_n(0)|^2 = 1$ (uniform distribution), while $x_n^0$ and $y_n^0$ are chosen to be different, which strongly affects the time of a soliton formation.

For the chosen length of a chain ($Ns = 51$) soliton becomes steady faster if

$$x_n^0 = |b_n(0)| \frac{(-1)^n}{\sqrt{2}}, \quad y_n^0 = |b_n(0)| \frac{(-1)^{n+1}}{\sqrt{2}}, \tag{42}$$

that is when $x_n^0$ and $y_n^0$ are <u>not</u> smooth functions of $n$ (for $\eta > 0$). (Figure 3.)
But if $x_n^0$ and $y_n^0$ are smooth functions of $n$ i.e. $x_n^0 = |b_n(0)|$, $y_n^0 = 0$, then a particle at first "spreads" over the chain and only after that starts concentrating into a soliton and, as a consequence, the time of the soliton formation grows considerably (Figure 4). See also online presentation graphics for Figure 3 and Figure 4 (**Presentation_1** in the Supplement, http://www.matbio.org/downloads_en/Lakhno_en2010(5_1)s.zip). For online demonstration we have chosen the sequence length to be $Ns = 71$.

### Delocalized initial state

Now let us consider how a soliton state is formed from a delocalized state of the form of (29). In Figs.5 and 6, the initial values of $|b_n(0)|$ are chosen in the form of:

$$|b_n(0)| = \frac{\sqrt{2}}{4}\sqrt{\frac{\varkappa\xi}{|\eta|}}\,\text{ch}^{-1}\left(\frac{\varkappa\xi(n-n_0)}{4|\eta|}\right), \quad \xi = 0.2, \quad n_0 = Ns/2. \tag{43}$$

The parameter $\xi = 0.2$ "stretches" the function of the inverse hyperbolic cosine over the whole length of the chain and we observe formation of a soliton state from a delocalized state of the form of (43). Here, as in Figs.3 and 4, the initial values of $x_n^0$ and $y_n^0$ are taken to be <u>un</u>smooth functions of $n$ of the form of (42) and smooth functions of $n$ such that $x_n^0 = |b_n(0)|$, $y_n^0 = 0$. In this case the difference in the times of a soliton formation for various choices of the initial $x_n^0$ and $y_n^0$ is much greater than in the case of a uniform initial distribution considered above. Besides, as compared to the uniform distribution, in the case of a smooth initial distribution function constructed with the use of an inverse hyperbolic cosine, the time of a soliton formation is much greater, and in the case of the <u>un</u>smooth distribution function (42) it is considerably less (for $\eta > 0$). In the same manner the time of a soliton formation depends on whether $x_n^0$ and $y_n^0$ are chosen to be smooth or <u>un</u>smooth functions, if the initial distribution $|b_n(0)|$ is chosen in the form of a "step" or a Gaussian function $|b_n(0)| = \Gamma e^{-g(n-n_0)^2}$.

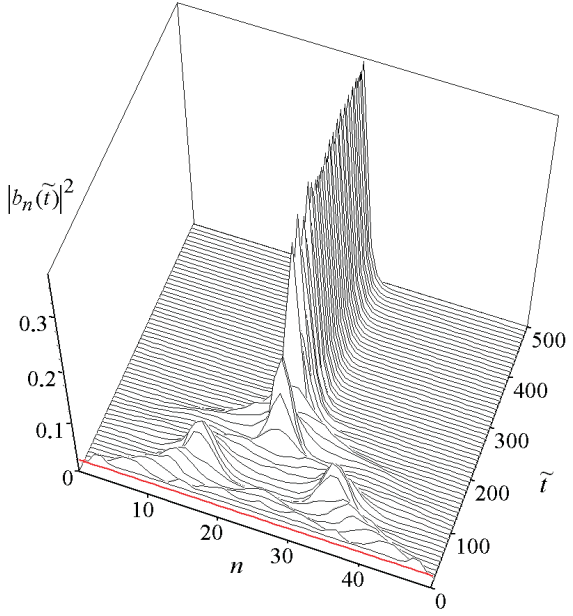
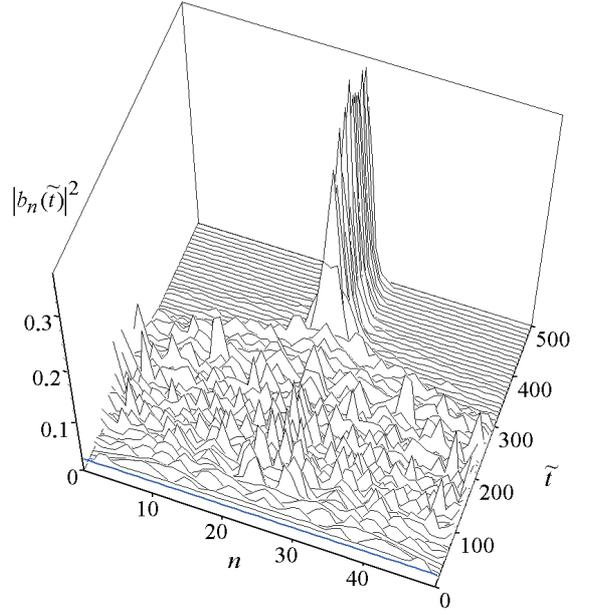

**Fig. 3.** Graphs of the functions $|b_n(t)|^2$ and $\delta(t)$, $\widetilde{E}$, $\widetilde{E}_Q$, $\widetilde{E}_{cl}$ for <u>un</u>smooth initial values; the initial distribution of a particle is uniform.

**Fig. 4.** Graphs of the functions $|b_n(t)|^2$ and $\delta(t)$, $\widetilde{E}$, $\widetilde{E}_Q$, $\widetilde{E}_{cl}$ for smooth initial values; the initial distribution of a particle is uniform.

### *Localized at one site*

A somewhat different picture arises if at the initial moment a particle is localized at one site: $x_n^0 = 0$ for $n \neq Ns/2$, $x_n^0 = 1$ for $n = Ns/2$. As can be seen from Figure 7, the distribution of a particle very quickly concentrates into a steady soliton. At the same time this picture slightly differs from that for the case of an <u>un</u>smooth initial distribution function constructed with the use of an inverse hyperbolic cosine, see Figure 5 and Figure 7.

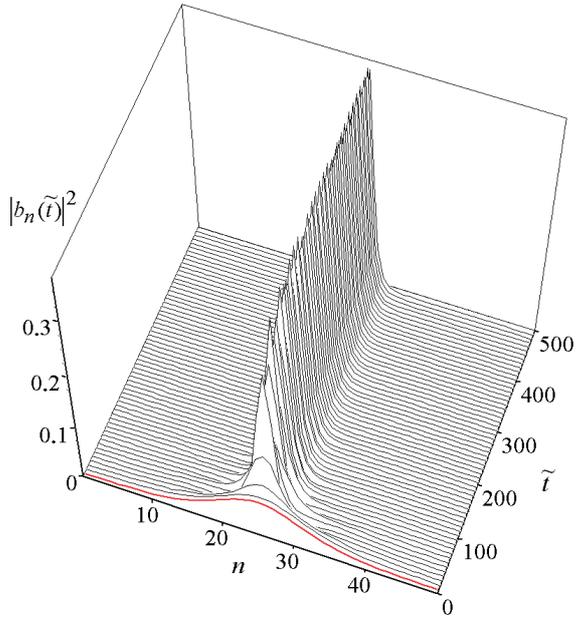

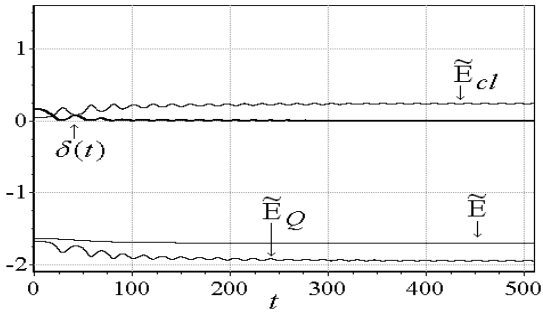

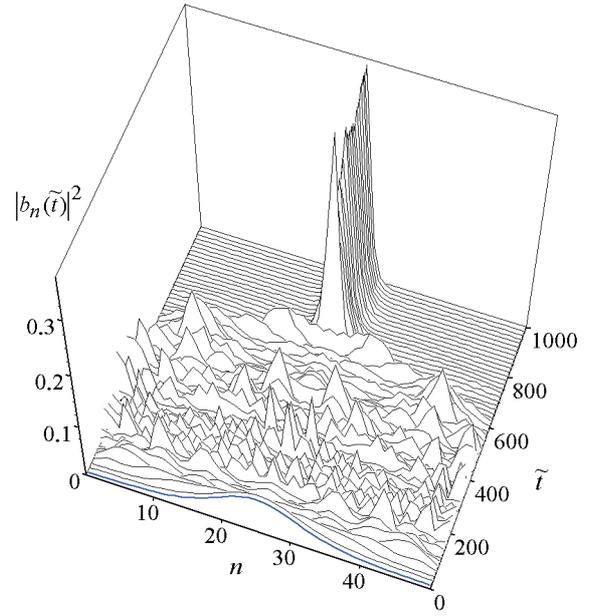

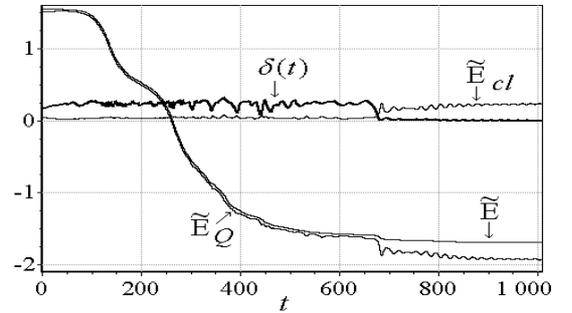

**Fig. 5.** Graphs of the functions $|b_n(t)|^2$ and $\delta(t)$, $\widetilde{E}$, $\widetilde{E}_Q$, $\widetilde{E}_{cl}$ for <u>un</u>smooth initial values (42); the initial distribution of a particle is constructed with the use of an inverse hyperbolic cosine (43).

**Fig. 6.** Graphs of the functions $|b_n(t)|^2$ and $\delta(t)$, $\widetilde{E}$, $\widetilde{E}_Q$, $\widetilde{E}_{cl}$ for smooth initial values; the initial distribution of a particle is constructed with the use of an inverse hyperbolic cosine (43).

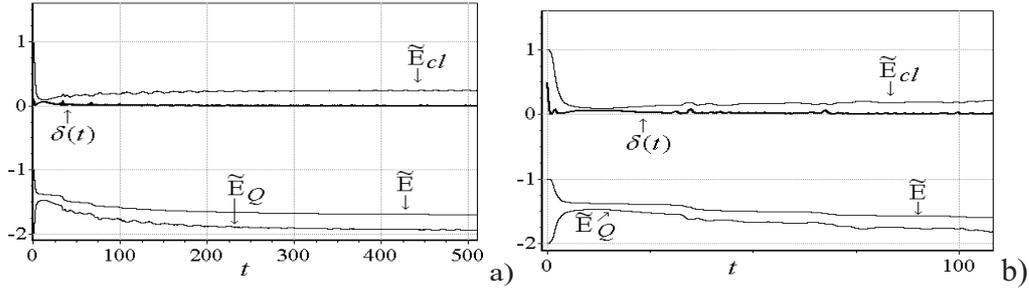

**Fig. 7.** Graphs of the functions $\delta(t)$, $\widetilde{E}$, $\widetilde{E}_Q$, $\widetilde{E}_{cl}$ for the following initial values: $y_n^0 = 0$, $x_n^0 = 0$ for $n \ne Ns/2$, $x_n^0 = 1$ for $n = Ns/2$, $u_n^0 = -|b_n(0)|^2/\omega_n^2$, a) on the time scale $t_{max} = 500$, b) on the time scale $t_{max} = 100$.

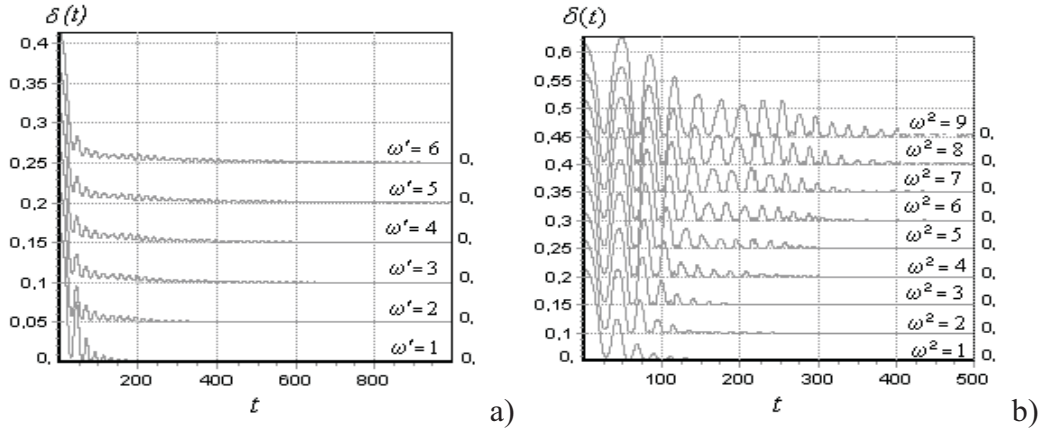

**Fig. 8.** Graphs of the function $\delta(t)$ displaced by 0.05 for: $\omega^2 = 1$, $\omega' = 1\ldots 6$ (a) and $\omega' = 1$, $\omega^2 = 1\ldots 9$ (b).

If $x_n^0 = -|b_n(0)|$, $y_n^0 = 0$ (smooth distribution), then on the graphs of the function $\delta(t)$ (Figure 4 and Figure 6), a large time takes the region, corresponding to "spreading" of a particle over the chain. By contrast, the graph $\delta(t)$ in Figure 5 suggests that in the case of an <u>un</u>smooth distribution, the function $\delta(t)$ immediately decreasing straight away, that is the particle immediately concentrating into a soliton. Therefore, to understand how the time of a soliton formation depends on the values of $\omega^2$ and $\omega'$, we take the initial values of $b_n(0)$, $x_n^0$ and $y_n^0$ identical to those in Figure 5.

Let us see how the time of a soliton formation depends on varying $\omega^2$ and $\omega'$ at fixed initial values of all other parameters. We will choose such $b_n(0)$, $x_n^0$ and $y_n^0$ for which a soliton becomes steady most quickly, namely:

- the initial distribution function $b_n(0)$ will be constructed with the use of the inverse hyperbolic cosine (43),
- the initial values of $x_n^0$ and $y_n^0$ will be chosen as <u>un</u>smooth functions of $n$, determined by relations (42),
- $u_n^0 = -|b_n(0)|^2/\omega_n^2$.

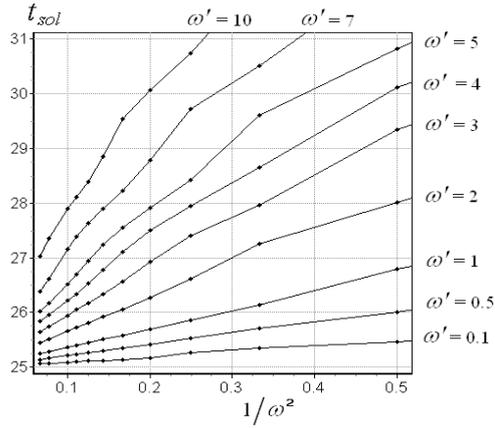
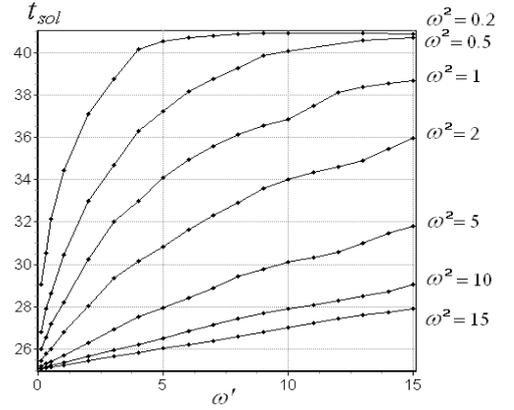

**Fig. 9.** Dependence of $t_{sol}$ on $1/\omega^2$ for various values of $\omega'$.

**Fig. 10.** Dependence of $t_{sol}$ on $\omega'$ for various values of $\omega^2$.

In the previous section we assessed the dependence of the soliton formation time on merely the initial values of $b_n$, $x_n$, $y_n$. To assess how this parameter depends on various values of $\omega^2$, $\omega'$ and $\omega^2/\omega'$ we will study the problem numerically. Let us fix $\omega^2 = 1$ and construct the graphs of $\delta(t)$ for various values of $\omega'$. In a similar manner we will construct the graphs of $\delta(t)$ for various values of $\omega^2$, at fixed $\omega' = 1$.

The graph in Figure 8.a suggests that the soliton formation time increases with growing $\omega'$, since at each instant of time the value of the mean-square deviation of $\delta(t) = \sum_{n=1}^{Ns}\left(|b_n(t)|^2 - |b_n(T_{sol})|^2\right)^2$ is the greater, the larger is the value of $\omega'$. At the same time the dependence of the soliton formation time on $\omega^2$ has a nonlinear character, see Figure 8.b. Analyzing the graphs of $\delta(t)$ constructed for various values of $\omega^2$ and $\omega'$ we can assess how the time of the soliton initial manifestation depends on $\omega^2$ and $\omega'$. Let us write $t_{sol}$ for the initial moment when a soliton becomes nearly steady, or, in other words, when the function $\delta(t)$ exhibits the first minimum.

Figure 9 illustrates the dependence of the time when a soliton becomes nearly steady $t_{sol}$ on the inverse value of the squared frequency $\omega^2$ for various fixed values of friction $\omega'$. It is easily seen that all the curves have regions of linear dependence, when the relation $\omega' \lesssim \omega$ is fulfilled. To analyze the dependence of the moment of the soliton first "manifestation" $t_{sol}$ on friction $\omega'$ for various fixed values of the squared frequency let us take advantage of Figure 10. We can see that the curves in Figure 10 also have regions of linear dependence at $\omega' \lesssim \omega$, besides, the dependence of $t_{sol}$ on $\omega'$ is nearly linear for the whole range of values, when $\omega$ is increasing. For example, for $\omega^2 = 15$ the dependence of $t_{sol}$ on $\omega'$ is linear for all the indicated values of $\omega'$. Analysis of Figure 9 and Figure 10 suggests that the larger is the value of $\omega'$, the greater is the time $t_{sol}$, and, the larger is the value of $\omega^2$, the less is the time $t_{sol}$. To put it another way, $t_{sol} \sim \omega'/\omega^2$, as is evident from expression (41).

## 8. FORMATION OF *N*-SOLITON LOCALIZED STATES IN UNIFORM CHAINS OF DIFFERENT LENGTHS

In the previous section we simulated formation of a stationary soliton state at such system's parameters when only one soliton is formed. We also considered how the dynamics of a steady soliton formation depends on the choice of various initial values of the functions $b_n(0)$, $x_n(0)$, $y_n(0)$ ($b_n = x_n + iy_n$). Since we deal with a symmetrical DNLS system, all the initial charge distributions chosen earlier are symmetrical, and for the chosen chains parameters only one soliton is formed, we do observe settling of this soliton in the center of the chain. If we assume that the initial distribution is uniform and extend the chain's length so that two solitons could be formed, these solitons will be symmetric about the chain center.

We will call such solutions multisoliton ones only by analogy with the one-soliton solution in the chain. Strictly speaking, many-peak formations considered in this section cannot be reckoned among solitons. Indeed, a real soliton must demonstrate asymptotic behavior with amplitude $b_n$ and displacement $u_n$ approaching zero as one moves farther and farther away from the soliton center. It is obvious that in a chain of a finite length, one cannot move away from the soliton center more than half this length and the role of boundaries is of importance. At the same time, if the distance between the solitons is much longer than the width of the soliton itself, the boundary conditions practically do not change the soliton characteristics and only determine its location in the chain. In this case for each individual peak-soliton of our multisoliton solution, asymptotic requirements imposed on a real soliton are fulfilled, namely $b_n \to 0, u_n \to 0$ as one moves farther and farther away from the center of each individual peak-soliton.

In this section we will consider formation of one-electron $N$-soliton localized states in uniform chains of various lengths. The number of arising peaks (solitons) depends on the chain parameters. In the case of a uniform initial distribution, as the chain's length increases (all the other parameters being fixed) we observe formation of various multisoliton distributions (see Figure 11). We may state that in any one chain of sufficient length, depending on the initial distribution, $N, N-1, \ldots, 1$ soliton states can be formed.

In Figure 11 for chains of various lengths, we show graphs of the functions $\left|b_n(T_{sol})\right|^2$ for rather long (for each graph) time $T_{sol}$ during which the multisoliton distribution becomes steady. The graphs presented were obtained from a uniform unsmooth initial distribution of the form of (42) for the following parameter values: $\varkappa = 4, \eta = 1.276, \omega = \omega' = 1$. We can see that the multisoliton distributions arising are symmetrical and in the case of sufficiently long chains consist of peaks different in height and width. We emphasize that for one and the same chain, the charge distribution can exhibit various peaks (equal also). Identical peaks are observed invariably only when a two-soliton distribution is formed. Breaking of the initial uniform <u>un</u>smooth (for $\eta > 0$) distribution starts from the edges of the chain and progresses to its center (see Figure 3 and any online presentation graphics in the Supplement, http://www.matbio.org/downloads_en/Lakhno_en2010(5_1)s.zip. Starting from the edges of the chain peaks arise one by one, their shape (height and width) depends on the chain's length (for fixed $\varkappa, \eta$), namely on the values of the amplitudes of the probabilities of the particle's occurrence on the chain sites: $|b_n(0)|^2 = 1/\sqrt{N}$. When the process of "failing" of the uniform distribution "reaches" the chain center, the process of formation of the maximum number of peaks still goes on. This time depends on the chain parameters, its length and the values of $\omega, \omega'$.

(See online presentation graphics **Presentation_2_a,_b** and **Presentation_3_a,_b** in the Supplement, http://www.matbio.org/downloads_en/Lakhno_en2010(5_1)s.zip). Via Presentation_2_b and Presentation_3_a,_b we can observe a simultaneous evolution of the functions $|b_n(t)|^2$ and $u_n(t)$ in the course of formation of a 2-soliton and 3-soliton localized state for different values of parameters. The less are the values of $\omega, \omega'$, the fewer "superfluous" peaks are formed. Then the process of merging the peaks starts and a steady state establishes. This process lasts longer for less values of $\omega, \omega'$. The number of peaks at the end of the process does not depend directly on the chain length. As the chain length increases, fewer peaks may form than in the case if it be slightly shorter, but when the chain length grows considerably, the number of peaks will always increase. The height of the first (from any side) peak is proportional to $|b_n(0)|^2 = 1/\sqrt{N}$.

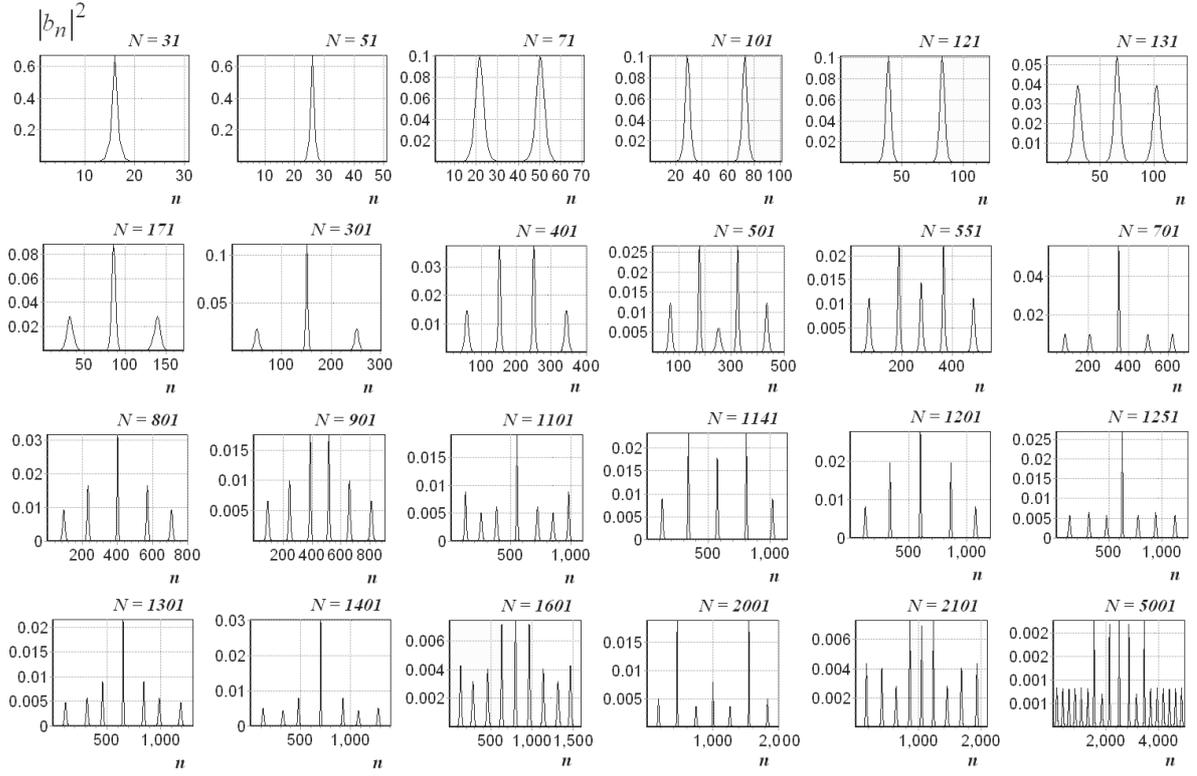

**Fig. 11.** Graphs of the function $|b_n(t)|^2$ for a steady distribution in the chains of different lengths. The initial charge distribution is uniform (<u>unsmooth</u>), the values of the chain parameters are the following: $\varkappa = 4, \eta = 1.276, \omega = \omega' = 1$. Some of the this graphics is better illustrated by the online presentation graphics for Figure 11, **Presentation_4_a,_b,_c** in the Supplement, http://www.matbio.org/downloads_en/Lakhno_en2010(5_1)s.zip

See also the online presentation graphics for Figure 11, **Presentation_4_a,_b,_c** in the Supplement, http://www.matbio.org/downloads_en/Lakhno_en2010(5_1)s.zip.

Besides, a multisoliton distribution can be obtained from a relevant initial (nonuniform) distribution. For example, we can take an initial distribution given by equations (44), (45), which determine an extended distribution close to the steady one:

$$b_p(n) = \sum_{j=0}^{p-1} b_n^j, \quad \left|\sum_{n=0}^{N-1} b_p(n)\right|^2 = 1, \qquad (44)$$

$$b_n^j = \xi_j \sqrt{\frac{\varkappa}{8|\eta|}} \cdot \text{ch}^{-1}\left[\frac{\varkappa \xi_j}{4\eta} \cdot (n - n_0^j)\right], \, j = 0, \ldots, p-1, \quad \sum_{j=0}^{p-1} \xi_j = 1 \qquad (45)$$

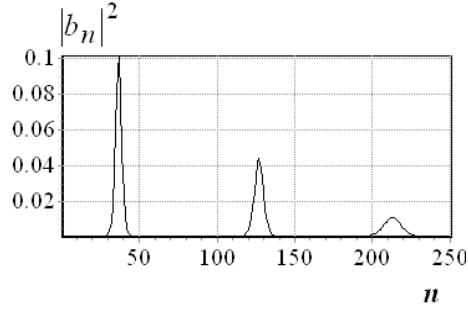

**Fig. 12.** Graph of the function $|b_n(t)|^2$ for a steady distribution obtained from the initial distribution of the form of (44), (45), where $p = 3$, $\xi_0 = 1/2$, $\xi_1 = 1/3$, $\xi_2 = 1/6$, and the following parameter values: $\varkappa = 4, \eta = 1.276, \omega = \omega' = 1$.

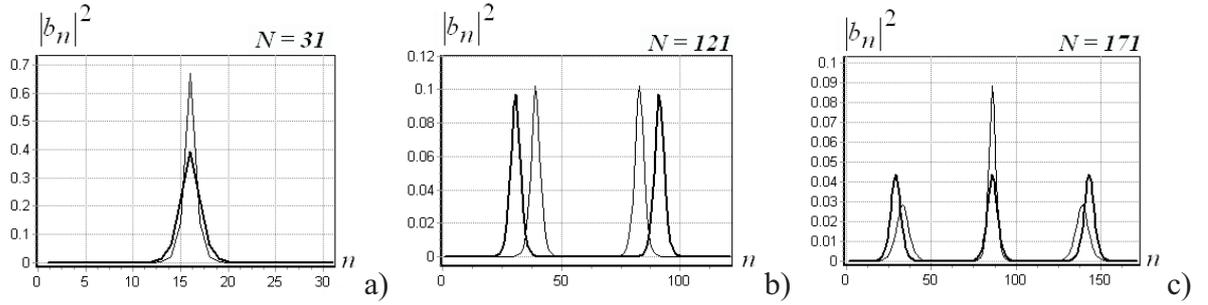

**Fig. 13.** Graphs of the functions $|b_n(t)|^2$ for a steady distribution obtained from a uniform <u>un</u>smooth initial distribution (thin line) and graphs of the functions $|b_p(n)|^2$ obtained in accordance with equation (47) for the cases $p = 1$, $p = 2$ $p = 3$ (thick line) for the following parameter values: $\varkappa = 4$, $\eta = 1.276$, $\omega = \omega' = 1$. This graphics is better illustrated by the online presentation graphics for Figure 13, **Presentation_5** in the Supplement, `http://www.matbio.org/downloads_en/Lakhno_en2010(5_1)s.zip`.

where $p$ is the number of preassigned peaks, $N$ is the number of sites in the chain, $n$ is the site's number, parameter $\xi_j$ spreads the function of each $j$-th inverse hyperbolic cosine "widthways" across the sites and proportionally decreases its height, $n_0^j$ are chosen depending on $\xi_j$ so that the normalizing condition

$$\left|\sum_{n=0}^{N-1} b_p(n)\right|^2 = 1 \qquad (46)$$

be fulfilled. Choosing $n_0^j$ one should bear in mind that the fewer is $\xi_j$, the greater is the spreading of the inverse hyperbolic cosine function across the sites, and hence, the peak with a less value of $\xi_j$ falls on a greater number of sites. It is obvious that the chain of a fixed length can house the maximum number of peaks when all the peaks are similar, i.e. all $\xi_j$ are equal.

The initial distribution corresponding to equations (44), (45) can be both symmetrical and asymmetrical. Figure 12 shows (as an example) a graph of the function $|b_n(t)|^2$ obtained from an asymmetrical initial distribution. For this graph, we chose the initial distribution $|b_n(0)|^2$ consisting of three ($p = 3$) peaks different in height and width, for which purpose we have set following values for $\xi_j$: $\xi_0 = 1/2$, $\xi_1 = 1/3$, $\xi_2 = 1/6$. For the chain length $N = 251$ we chose the values $n_0^0 = 41, n_0^1 = 126, n_0^2 = 211$, for which the normalizing condition (46) is fulfilled with sufficient accuracy ($1 \pm 10^{-5}$). Figure 12 shows only a graph of the function $|b_n(t)|^2$ for the sufficiently large time, i.e. this is the graph of a steady distribution, since the graph of the chosen initial function $|b_n(0)|^2$ practically coincides with it. The asymmetric shape of the charge distribution is retained for a very long time, calculations were carried out for the time $T > 3 \cdot 10^7$.

Solution of equations (3), (4) in the continuum approximation leads to the following approximate function $b_p(n)$ in the discrete chain [31, 32, 33]:

$$b_p(n) = \frac{1}{p} \cdot \sqrt{\frac{\varkappa}{8|\eta|}} \cdot \sum_{j=0}^{p-1} (-1)^j \cdot \text{ch}^{-1}\left[\frac{\varkappa}{4\eta p^2} \cdot \left(pn + N\left(\frac{p-1}{2} - j\right)\right)\right]. \quad (47)$$

This function determines a near steady distribution which consists of $p$ equal peaks equidistant from one another. As in the case of the distribution determined by equation (45), the normalizing condition (46) should be fulfilled. The maximum number of peaks that the finite-length chain can house is determined (or limited) by the fulfillment of the normalizing condition (46) with a required accuracy. If we take the initial distribution given by equation (47), the steady distribution (for $\omega' \neq 0$) will be practically the same as the initial one, and only for $p = 1$ the steady soliton is much higher and slightly thinner than the initial one (see Figure 13 a)). Notice also that the initial position of the peaks in the chain remains unchanged for any number of peaks, as well as for $p = 1$, in other words, all the peaks remain motionless with respect to $n$ (for $\omega' \neq 0$). In order for the peaks of the predetermined initial distribution to remain motionless in the absence of friction, $\omega' = 0$ and $p > 1$, the initial distribution given by equation (47) should be set with multiple precision of the fulfillment of the normalizing condition (46) and the chain length should be an order of magnitude larger than in the case of $\omega' \neq 0$. The reason is that the influence of the chain boundaries on the charge distribution and the influence of neighboring peaks against each other in a discrete system (especially at large $\varkappa$) is not compensated for by the availability of even small friction.

Figure 13 compares graphs of the functions $|b_n(t)|^2$ for a steady distribution obtained from a uniform <u>unsmooth</u> initial distribution (thin line) and graphs of the functions $|b_p(n)|^2$ obtained in accordance with equation (47) for $p = 1$, $p = 2$ and $p = 3$ (thick line).

Notice, that the total energy corresponding to multisoliton state (47) has the form:

$$\widetilde{\text{E}}_{cont}^p = -\frac{\varkappa^2}{48\eta p^2}. \quad (48)$$

Table 1 and Fig.14 compare the values of expression $\widetilde{\text{E}}_{cont}^p$ (48) (continuum approximation) with the value of the total energy obtained numerically (in a discrete model). We can see that the difference between the theoretical and numerical values decreases by an order of magnitude as $p$ increases by 1. As the table suggests, a good agreement between the theoretical values of $\text{E}^p$ and the numerical ones is observed for $p > 1$, since in this case continuum approximation is fulfilled much better, than in the case of $p = 1$. Notice also that as $\varkappa$ decreases ($\eta$ being fixed), continuum approximation is fulfilled better. In the foregoing we mentioned that if we take the initial distribution determined by equation (47), the steady-state distribution (for $\omega' \neq 0$) will be practically the same as the initial one, and only for the case of $p = 1$ the steady soliton is much higher and thinner than the initial one (see Figure 13 a)).

**Table 1.** Comparison of the values of the total energy $\widetilde{E}^p_{cont}$ obtained from (48) (continuum approximation) with numerical results calculated for the total energy $\widetilde{E}^p_{discr}$ (in a discrete model) for different numbers of peaks in $N$ - soliton distribution for the following parameter values: $\varkappa = 4$, $\eta = 1.276$, $\omega = \omega' = 1$. Values of dimensionless quantities $\widetilde{E}^p$ are related to their values in $eV$ by $E^p(eV) = \widetilde{E}^p/15.19$

| $p$ | $\widetilde{E}^p_{cont}$ | $\left(\left\|\widetilde{E}^p_{cont} - (\widetilde{E}^p_{discr} + 2\eta)\right\|\right)_{at\ t=0}$ | $\left(\widetilde{E}^p_{discr} + 2\eta\right)_{at\ t=0}$ | $\left(\widetilde{E}^p_{discr} + 2\eta\right)_{steady}$ |
|---|---|---|---|---|
| 1 | - 0.261233 | 0.019103 | - 0.280336 | - 0.317596 |
| 2 | - 0.065308 | 0.001152 | - 0.066460 | - 0.066569 |
| 3 | - 0.029026 | 0.000229 | - 0.028334 | - 0.029224 |
| 4 | - 0.016327 | 0.000073 | - 0.015301 | - 0.016394 |

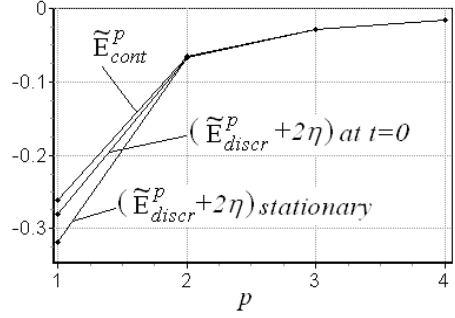

**Fig. 14.** Graphical comparison of the value of expression $\widetilde{E}^p_{cont}$ (48) (continuum approximation) with the value of the total energy $\widetilde{E}^p_{discr}$ found numerically, $p$ - is the number of peaks in the chain. Numerical values for these graphs are presented in Table 1.

This fact is also illustrated in Table 1 and in Fig.14 which present numerical values and graphics of the total energy $\widetilde{E}^p_{discr} + 2\eta$ in a discrete system for the distribution of the form of (47) at the initial moment of time and for a steady distribution at the sufficiently large time.

In conclusion it may be said that in view of condition (46), the multisoliton structures under consideration represent formations with a fraction electron charge. According to (47), each peak (soliton) determined by this relation contains a charge of value $e/p$. Such a fraction charge can be measured experimentally since its availability is detected by a classical device - deformation of a chain having a peak distribution (Figure 15) in the region of each peak.

## 9. FORMATION OF A SOLITON IN A UNIFORM POLYNUCLEOTIDE CHAIN

*DNA* holds a most unique position among molecules. A *DNA* molecule resembles a solid body. Base-pairs are stacked there in the same manner as in a crystal. But this is a linear, so to say, a one-dimensional crystal - each base-pair has only two nearest neighbors. Recall that *DNA* consists of four types of nucleotides denoted as *A* (adenine), *T* (thymine), $C$ (cytosine), $G$ (guanine) which form complementary pairs: nucleotide $A$ always pairs with $T$ and nucleotide $C$ pairs with $G$. These nucleotide pairs are arranged as a stack to form a $DNA$ double helix. At present long sequences with a predetermined sequence of nucleotide pairs can be synthesized artificially. Of interest are chains composed of similar pairs since they can be used as molecular wires in nanoelectronic devices [1, 5]. In the majority of experiments on charge transfer in $DNA$, charge carriers are holes rather than electrons. If we remove one electron from any nucleotide of the chain, the hole that will appear will have the potential energy $U < U_G < U_A < U_C < U_T$. Overlapping of electron $\pi$-orbitals of neighboring nucleotide pairs will lead to delocalization of the hole over the chain and its trapping by nucleotides with lower oxidation potential. Since, in accordance with the inequalities presented, guanine has the lowest oxidation potential, the hole will hop over guanines while all the other nucleotides will be potential barriers for its motion.

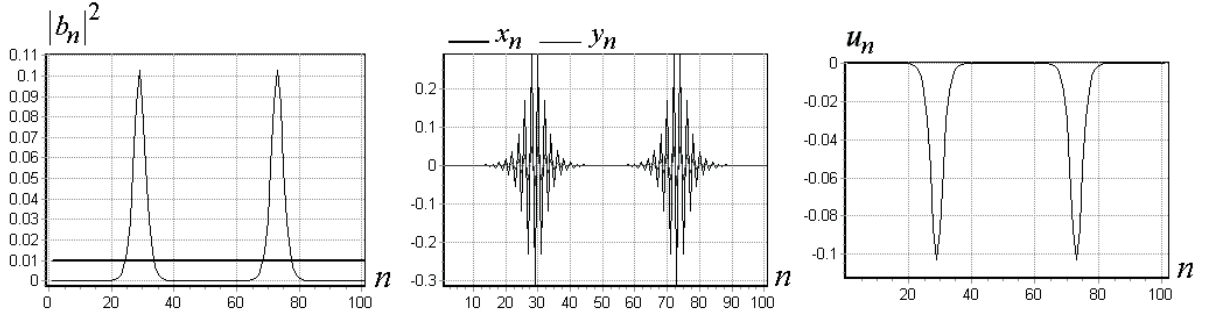

**Fig. 15.** Graphs of the functions $|b_n|^2$ ($b_n = x_n + iy_n$), $x_n$, $y_n$, $u_n$ for a steady distribution obtained from a uniform unsmooth initial distribution $|b_n(0)|^2$ (thick line) for the case $p = 2$ and for the following parameter values $\varkappa = 4$, $\eta = 1.276$, $\omega = \omega' = 1$. The length of a chain is equal 101.

Here we deal with a polynucleotide chain which differs from the model molecular chain considered above in the values of the parameters. If, in modelling the dynamics of a homogeneous $(G/C)_n$ nucleotide chain in which a charge is carried by holes, we choose the parameter values used in [9], namely $\eta = 1.276$, $\omega = 0.01$, $\omega' = 0.006$, $\varkappa = 4$, which correspond to physical parameters: $\nu = 0.084 eV$, $\Omega = \sqrt{k/M} = 10^{12} sec^{-1}$, $\Omega' = \gamma/M = 6 \cdot 10^{11} sec^{-1}$, $M = 10^{-21} g$, $\alpha' = 0.13 eV$ (which is close to the estimate obtained in [34]), then inequality (24) will not be fulfilled and, accordingly, the estimate (41) will be inapplicable. In this case the picture of a soliton formation differs from that described in section 7.

In a $(G/C)_n$ nucleotide chain with the above-indicated parameters, as in the general case of a molecular chain, at an arbitrary initial density distribution $|b_n(0)|^2$, the evolution process can be divided into three phases, if the chain was not initially deformed: $(u_n(0) = 0)$.

In the first phase a particle "spreads" over the chain demonstrating long-term quasichaotical density oscillations $|b_n(t)|^2$.

In the second phase a delocalized state of the form of (43) is formed from the chaotic stage with uniform mead distribution of $|b_n(t)|^2$ on a chain sites. The second stage deals with oscillationless deformation of the squared modulus of the delocalized state wave function (43) accompanied by the formation of a potential well, caused by the displacement $u_n$, which culminates in the formation of a localized dynamical soliton.

The third phase is the oscillating stage of the dynamical soliton state which at the parameter values concerned does not evolve into a steady soliton during the time of the calculations.

In the Appendix we show that the first phase of the particle "spreading" over the chain is unstable relative to the formation of a localized state.

First, let us consider the formation of a soliton at $\eta = 1.276$, $\varkappa = 4$ ($\eta$, $\varkappa$ are the same as for DNA), but for large values of $\omega = \omega' = 1$. Here we observe a quick formation of a steady soliton (see Figure 13a)) for any initial distribution of a particle. In this case the graphs of the functions $|b_n(t)|^2$ and $u_n(t)$ become motionless and the function values get invariant, and $du_n/dt$ vanish. Therefore this state can be considered stationary. The energy values for this state are the following: $E = -2.869$, $E_Q = -3.848$, $E_{cl} = 0.978$.

Now let us consider the formation of a soliton in a homogeneous $(G/C)_n$ nucleotide chain with the parameters identical to those of DNA, namely $\eta = 1.276$, $\varkappa = 4$, $\omega = 0.01$, $\omega' = 0.006$ for various initial distributions of a charge.

### *Initial distribution is an analytical inverse hyperbolic cosine*

If the initial distribution of a particle is constructed from an analytical inverse hyperbolic cosine of the form of (43) at $\xi = 1$, then, it would seem, we observe the formation of a steady soliton for chains of various length. The graphs of the functions $|b_n(t)|^2$ and $u_n(t)$ become almost motionless and the values of this functions - nearly constant, but $du_n/dt$ still have nonzero values, small as they are ($\approx 0.001$) for rather a long time. In this case the energy values differ slightly from the above-considered stationary state: $E = -2.864$, $E_Q = -3.834$, $E_{cl} = 0.969$, and the total energy E is a little larger, than that in the stationary state described above. It is just this not full relaxing which is responsible for nonzero values of $du_n/dt$.

But if the initial distribution is uniform, localized at one site, or constructed with the use of a stretched inverse hyperbolic cosine, we no longer observe the formation of a steady soliton (for the indicated values of $\omega$ and $\omega'$) and the graph of the function $\delta(t)$, representing a mean-square deviation of $|b_n(t)|^2$ from a "steady" soliton, does not tend to zero during the time of the calculations: namely $\delta(t) \gtrsim 0.05$. For all considered cases the graphs of the functions $|b_n(t)|^2$ and $u_n(t)$ are in a slight motion and the values of $du_n/dt$ hold large ($\approx 0.015, 0.04$) for a long time $T > 3 \cdot 10^6$.

### *Initial distribution is uniform or localized at one site*

If the initial distribution is uniform or localized at one site (and $n = 51$), the function $|b_n(t)|^2$ oscillates near an analytical soliton in the center part, and is different from zero demonstrating quasichaotic oscillations outside the soliton region where a soliton is characterized by zero values of $|b_n(t)|^2$. For other values of $n$ the graphs of the function $|b_n(t)|^2$ can be "above" or "below" the analytical soliton. As $n$ increases, two or more solitons are formed (in the case of a uniform initial distribution).

### *Initial distribution is a stretched inverse hyperbolic cosine*

The distribution function demonstrates a somewhat different behavior, if the initial distribution is constructed with the use of a stretched inverse hyperbolic cosine. The graphs of $|b_n(t)|^2$, presented in Figure 16 suggest that if the initial distribution is constructed from an inverse hyperbolic cosine of the form of (43) stretched over the whole length of the chain at $\xi = 5$, the graph of $|b_n(t)|^2$ ($n = 51$) does not tend to take the shape of a graph for a steady soliton, but "fits" into the graph of an analytical hyperbolic cosine nearly coinciding with it in height. However it does not faithfully copies its shape, being narrower in the central part and having nonzero values outside it.

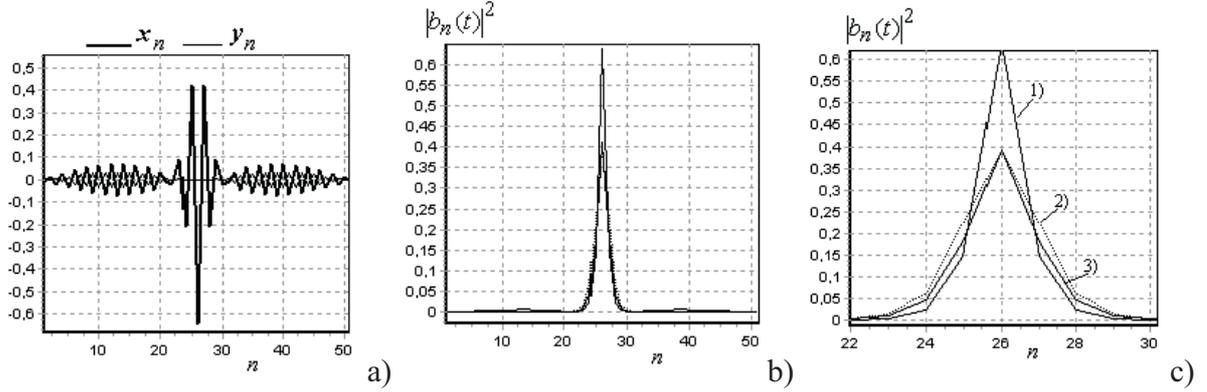

**Fig. 16.** Graphs of the functions a): $x_n(t), y_n(t)$ and b),c): $|b_n(t)|^2$ for <u>un</u>smooth initial values (32), the initial distribution of a particle is constructed with the use of the inverse hyperbolic cosine (43) at $\xi = 5, n = 51$. Figure 12.c) presents the central part of Figure 12.b). 1)— graph of the function $|b_n(t)|^2$ for a steady soliton (——), 2)— graph of the inverse hyperbolic cosine (43) at $\xi = 1, \varkappa = 4, \eta = 1.276$ (······), 3)— graph of the function $|b_n(t)|^2$ for $T > 1000$ and for $T > 1000000$ (——).

Besides, as distinct from the above-considered cases, oscillations of the distribution function are very slight in the central part and the function fits into the analytical soliton almost immovably, while outside the soliton region we observe not chaoticall oscillations but also a nearly motionless graph of the function $|b_n(t)|^2$, see Figure 16. This shape of the charge distribution curve holds for rather a long time, calculations were performed for $T$ exceeding $6 \cdot 10^6$. The energy values at the moment $T = 6 \cdot 10^6$ are the following: $E = -2.7308$, $E_Q = -3.169$, $E_{cl} = 0.4386$.

If the initial distribution of a particle is constructed from the stretched inverse hyperbolic cosine (43) for various values of $\xi, n$, then the graph of the function $|b_n(t)|^2$ does not tend to take the shape of a graph for a steady or analytical soliton at large $T$. In the central part the graph of $|b_n(t)|^2$ can run above the graph for the analytical soliton and be narrower, or else can run below it and be wider, depending on the values of $\xi$ and $n$. Outside the central part, the function $|b_n(t)|^2$ has nonzero values and non-mobile graph.

Let us discuss the periods observed in the oscillations of the charge density distribution. Let us consider the final, oscillating stage of the soliton dynamical state. The distribution function $|b_n(t)|^2$, displacements $u_n(t)$ and $du_n/dt$ have pronounced oscillation periods. If the initial distribution is taken to be uniform, localized at one site or constructed with the use of an analytical inverse hyperbolic cosine, then the oscillation periods of the above-indicated functions are small ($\approx 2 - 4$) and the oscillations are irregular, rather chaotical and low-amplitude.

But if the initial distribution of a particle is constructed from a stretched inverse hyperbolic cosine, the oscillation periods of the above-indicated functions have a pronounced dependence on the parameters $\eta$ and $\varkappa$. Since the oscillation periods of the functions $|b_n(t)|^2$ and $u_n(t)$ are similar, we will consider the dependence of the oscillation periods of the functions $|b_n(t)|^2$ in the center of the molecular chain, see Figure 17. Oscillations of the distribution function in the central part differ from those outside the soliton region. Still more pronounced are the differences in the oscillations of the functions $du_n/dt$ in the center of the chain and outside. In the center of the chain oscillations of the function $du_n/dt$ are approximately 11 times more frequent.

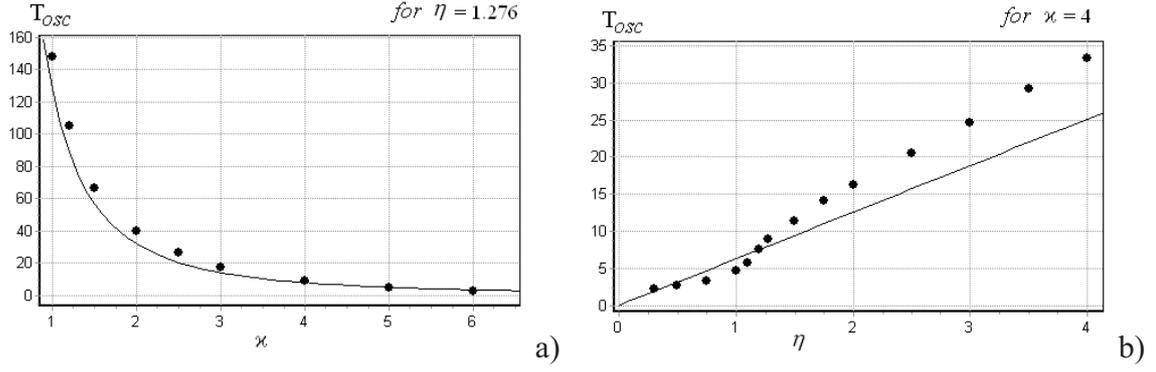

**Fig. 17.** Bold dots (•) indicate the dependence of the oscillation period of the function $|b_n(t)|^2$ at the point $n = Ns/2 + 1$ at large values of $t$ ($t > 10^6$), a)— for various values of $\xi$ at $\eta = 1.276$, b)— for various values of $\eta$ at $\xi = 4$. The solid line presents an analytical approximation of the oscillation period, i. e. the graph of the function $T_{osc} = 32\pi \frac{\eta}{\varkappa^2}$: a)— for $\eta = 1.276$, b)— for $\varkappa = 4$.

## 10. DISCUSSION OF THE RESULTS

The dependence of the soliton formation time on the initial phase of the wave function $b_n$ may appear strange at first sight, since the classical motion equations for sites (4) involve a squared modulus of the wave function $|b_n|^2$. For this reason at the initial moment, the evolution of the site displacements from their equilibrium positions is the same for any initial values of the amplitude phases. However, this statement is not valid for the quantum-mechanical part of the problem (3). In the case under consideration the dependence of the soliton formation time on the initial phase is concerned with the following fact.

In our numerical experiments the value of the matrix element $\nu$ was chosen to be positive: $\eta > 0$. At the same time, if $b_n$ is an oscillating function, we can pass on, with the use of (9), to the smooth function $\tilde{b}_n$ replacing $\nu$ by $-\nu$. This transformation, as was shown in section 2, yields the energy value for $\tilde{b}_n$ identical to that for $b_n$. In the case of the function $\tilde{b}_n$ this energy is determined by the functional $E_{-\nu}[\tilde{b}_n]$ (28). In the case of the unsmooth initial wave function chosen in section 7, $E_{-\nu}[\tilde{b}_n(0)] < 0$, and the functional is limited from below. In this case the evolution proceeds in the direction of a smooth decrease in the initial distribution characteristic size and formation of a localized soliton state at which the total energy functional $E_{-\nu}[\tilde{b}_n]$ is minimum.

If the initial wave function $b_n$ is chosen to be smooth, the functional $E_{-\nu}[\tilde{b}_n(0)]$, which is the total initial energy, is positive. This initial wave function is unstable relative to "spreading" over the chain. In this case the localization process is preceded by an extended phase of the "spread out" state and only after that a particle becomes localized. For this reason, the soliton formation time turns out to be much greater than in the previous case. If $\nu < 0$, the situation is quite the opposite: the soliton formation time is less in the case of a smooth initial wave function, than in the case of an unsmooth one.

The results obtained demonstrate important distinctions between "discrete" descriptions of quantum systems and their "continuous" analogs. The discrete quantum-classical model considered by us, in which the continuous spatial coordinate is changed for a discrete lattice with nodes-sites is more symmetrical than the continuous model corresponding to it. In particular, functional (18), obtained from a discrete model symmetrical about the transformation $b_n \to \tilde{b}_n \exp i\pi n$, $\nu \to -\nu$, as a result of a limit transition, looses this symmetry in the continuum approximation. In passing on to a continuous model, the symmetry is lost at the stage of discarding the solutions which quickly oscillate in spatial variables and therefore cannot be described in the continuum approximation.

Notwithstanding these difficulties, a change-over from a discrete model to a continuous one enables one to obtain some informative results where this change-over is valid. This is perfectly illustrated by the dependence of the soliton formation time obtained for various parameter values in the course of numerical experiments with a discrete chain in comparison with the results obtained with the use of a continuum model (Figures 8,9).

The time of a soliton formation in DNA obtained in section 9 can be used for the analysis of recent experiments on charge transfer in uniform $(G/C)_n$ polynucleotide chains. The work [35] dealt with the charge transport in a $(G/C)_n$ chain with $n = 30$. According to [35], the maximum transfer rate in the chain of 30 nucleotide pairs (which corresponds to the chain of length 10.4 nm), was as great as $10^{12}$ charges/sec (i.e. $\sim 100n\text{Å}$). This corresponds to the maximum time of the charge occurrence in the chain $\tau \approx 30$ps. According to the results obtained (section 9) this time is sufficient for a soliton to form. In [35] the current value did not exceed $1n\text{Å}$, which corresponds to the time of the charge occurrence in DNA equal to $3 \cdot 10^3$ps. This time exceeds the time required for a soliton to form by more than two orders of magnitude.

So, the results obtained testify to the possibility of a soliton mechanism of the charge transfer in experiments [35].

Presently there is a large number of the works, devoted to the soliton state properties in molecular chains which have been initiated by Davydov's works, etc. [36, 37]. The interesting dynamic phenomena arise, when in a chain coexist any soliton states (see for example [38] and the literature cited therein).

Notice, that the estimate of the soliton formation time given by (40),(41) is valid on the assumption that in the system under consideration there is only one steady soliton state in which the evolution just culminates. This assumption is valid for not-too-long chains, such as those with $N = 51$ used by us. In longer chains this condition is violated and in the course of evolution, the initial delocalized state can turn into steady states of a different type, which could, by analogy with the classical case, be called quantum-classical dissipative structures.

The states of $p$-soliton form considered in section 8 can be used for recording information in nucleotide chains by electron rather than chemical technique. In this case each peak carries information about its fraction charge of value $\approx e/p$, where $e$ is the electron charge. Limitation on the number of peaks and the distance between them results from strict quantum-mechanical consideration in which oscillations of the chain sites are considered quantum-mechanically [39]. In this case exact solution of the problem is possible only in the limit cases of weak and strong interaction of an electron with the sites oscillations. According to [39], in the stationary state a semiclassical limit is achieved on condition that $\text{E}^p > \Delta\text{E}$, where $\text{E}^p$ is the total energy of the $p$-soliton state (48), $\Delta\text{E}$ is the electron energy in the limit of weak interaction of the charge with the chain oscillations. For $\Delta\text{E}$, in [39] an expression $\Delta\text{E} = \sqrt{\mu a^2 (\alpha')^4 / 8M^2 \hbar \omega^3}$, $\mu = \hbar/2\nu a^2$ which in dimensionless variables (5) is written as $\Delta E = \sqrt{\varkappa^2 \widetilde{\omega}/16\eta}$ was obtained. Accordingly, $E^p = \varkappa^2/48\eta p^2$. The requirement $E^p > \Delta E$ leads to the condition $(\varkappa/16)\sqrt{1/\eta\widetilde{\omega}} > p$. For the parameters of the *PolyG/PolyC* chain this gives: $2.2 > p$. So, in polynucleotide chains, both one- and two-soliton states of an electron are possible. This conclusion can also be made from qualitative considerations. Equation (47) suggests that the characteristic size of an electron state in an individual soliton is equal to the soliton's characteristic size: $r = 4\eta p a/\varkappa$. Then the energy of the particle localized in the region $r$ is $W_p = \hbar^2/2mr^2 = (\varkappa^2/16\eta p^2) \cdot \hbar/\tau$. The adiabaticity condition $W_p \gg \hbar\omega$ leads to the estimates obtained above.

In some works [40] is discussed a possibility to move individual peaks arbitrarily far apart from one another and thus obtain a formation with a fraction charge. The above estimates demonstrate inconsistency of such ideas. A complete quantum-mechanical description (when the chain is also described quantum-mechanically [39]) gives a picture which suggests that when the peaks are separated by rather a large distance, semiclassical description becomes inapplicable, solitons destroy and evolve into plane waves. In view of instability of plane waves (see Appendix) one-, two-, etc. soliton states will again arise in the chain. Nevertheless, under conditions of the model applicability, when all the mentioned conditions are met, the states with fraction charge may be observed experimentally, for example, upon application of an external field which causes motion of the solitons. In this case charge carriers will be particles with fraction charge.

The work was done with the support from the RFBR, project 07-07-00313.
The authors are thankful to the Joint Supercomputer Center of the Russian Academy of Sciences, Moscow, Russia for the provided computational resources.

## APPENDIX
## STABILITY OF A UNIFORM DISTRIBUTION
## OVER A MOLECULAR CHAIN

In section 6 we showed that uniform distribution of a charge over the chain corresponding to the case $\xi = 0$ is unstable with respect to formation of a soliton state since the total energy $E_v(t)$ decreases as $\xi$ grows. This conclusion does not depend on whether condition (24) is fulfilled or not. According to the results of section 6, when condition (24) is fulfilled, the soliton formation process is completely relaxed. When condition (24) is not fulfilled, the soliton formation process has dynamical unrelaxed character and instability of a uniform distribution of a charge over the chain is caused by instability of oscillating excitations of a uniform distribution.

The overall consideration of the arising of instability is based on the analysis of a continuous analog of the system of equation (6), (7):

$$i\frac{\partial b(n,\tilde{t})}{\partial \tilde{t}} = \eta \frac{\partial^2 b(n,\tilde{t})}{\partial^2 n} + \varkappa \omega^2 u(n,\tilde{t}) b(n,\tilde{t}), \qquad (49)$$

$$\frac{\partial^2 u(n,\tilde{t})}{\partial \tilde{t}^2} = -\omega' \frac{\partial u(n,\tilde{t})}{\partial \tilde{t}} - \omega^2 u(n,\tilde{t}) - |b(n,\tilde{t})|^2. \qquad (50)$$

We will seek solutions of (49), (50) excited with respect to the uniform state in the form:

$$b(n,\tilde{t}) = [1 + b_1(n,\tilde{t})]b_0(n,\tilde{t}), \ u(n,\tilde{t}) = u_0(n,\tilde{t})(1 + \varphi(n,\tilde{t})), \qquad (51)$$

where:
$$b_0(n,\tilde{t}) = a_0 e^{i\varkappa|a_0|^2 t} \quad (52)$$

is a uniform distribution of a particle over the chain with the same probability of distribution $|a_0|^2$ at each site of the chain, $u_0 = |a_0|^2/\omega^2$ is the displacement of sites corresponding to this uniform distribution. On the assumption that $b_1(n,\tilde{t})$ and $\varphi(\tilde{t})$ are small excitations of the uniform distribution, substitution of (51),(52) into (49),(50) yields for $b_1(n,\tilde{t})$ and $\varphi(\tilde{t})$ the following equations:

$$i\frac{\partial b_1(n,\tilde{t})}{\partial \tilde{t}} = \eta\frac{\partial^2 b_1(n,\tilde{t})}{\partial^2 n} - \varkappa|a_0|^2\varphi(\tilde{t}), \quad (53)$$

$$\frac{\partial^2 \varphi(\tilde{t})}{\partial \tilde{t}^2} = -\omega'\frac{\partial \varphi(\tilde{t})}{\partial \tilde{t}} - \omega^2\varphi(\tilde{t}) + \omega^2(b_1(n,\tilde{t}) + b_1^*(n,\tilde{t})). \quad (54)$$

We will seek the solutions of the linearized system (53),(54) in the form:

$$b_1(n,\tilde{t}) = c_1 e^{i(kn+\Omega\tilde{t})} + c_2 e^{-i(kn+\Omega\tilde{t})}, \quad (55)$$

$$\varphi(\tilde{t}) = \varphi_1 e^{i(kn+\Omega\tilde{t})} + \varphi_2 e^{-i(kn+\Omega\tilde{t})}. \quad (56)$$

Let us consider the case $\omega' \neq 0$.

Substitution of (55),(56) into (53),(54) (for $\omega' \neq 0$) leads to the dispersion equation of the form:

$$\Omega^6 - \Omega^5 \cdot 2i\omega' - \Omega^4(2\omega^2 + k^4\eta^2 + (\omega')^2) + \Omega^3(2i\omega^2\omega' + 2ik^4\eta^2\omega')$$
$$+\Omega^2(\omega^4 + 2k^4\omega^2\eta^2 + 2k^2\omega^2\eta\varkappa|a_0|^2 + k^4\eta^2(\omega')^2) \quad (57)$$
$$-\Omega(2ik^4\omega^2\eta^2\omega' + 2ik^2\omega^2\eta\varkappa|a_0|^2\omega') - k^4\omega^4\eta^2 - 2k^2\omega^4\eta\varkappa|a_0|^2 = 0$$

If there exist $k$, at which equation (57) contains complex conjugate values of $\Omega$ as a solution, then, according to (55),(56), this means that the solutions corresponding to a uniform distribution are unstable relative to long-wave excitations of this distribution.
Let us rewrite equation (57) as:

$$(\omega^2 - \Omega^2 + i\Omega\omega')(2k^2\omega^2\eta\varkappa|a_0|^2 + (k^4\eta^2 - \Omega^2)(\omega^2 - \Omega^2 + i\Omega\omega')) = 0 \quad (58)$$

Two simplest (by sight) roots of this equation are:

$$\Omega_1 = \frac{1}{2}\left(i\omega' - \sqrt{4\omega^2 - (\omega')^2}\right), \quad \Omega_2 = \frac{1}{2}\left(i\omega' + \sqrt{4\omega^2 - (\omega')^2}\right). \quad (59)$$

Therefore, in the case of $\omega' \neq 0$ dispersion equation (57) always contains complex conjugate roots at any parameter values which means that a uniform distribution is always unstable.

Consider next the case $\omega' = 0$.

For the case $\omega' = 0$ we get a dispersion equation of the form:

$$\Omega^6 - \Omega^4(2\omega^2 + k^4\eta^2) + \Omega^2(\omega^4 + 2k^4\omega^2\eta^2 + 2k^2\omega^2\eta\varkappa|a_0|^2) - k^4\omega^4\eta^2 - 2k^2\omega^4\eta\varkappa|a_0|^2 = 0. (60)$$

The roots of equation (60):

$$\Omega_{1,2} = \pm\omega, \quad \Omega_{3,4,5,6} = \frac{1}{\sqrt{2}}\sqrt{\omega^2 + k^4\eta^2 \pm \sqrt{\omega^4 - 2k^4\omega^2\eta^2 + k^8\eta^4 - 8k^2\omega^2\eta\varkappa|a_0|^2}},$$

can be complex if any expression under the radical sign in $\Omega_{3,4,5,6}$ is negative:

$$p = \omega^2 + k^4\eta^2 \pm \sqrt{\omega^4 - 2k^4\omega^2\eta^2 + k^8\eta^4 - 8k^2\omega^2\eta\varkappa|a_0|^2} < 0, \tag{61}$$

or

$$q = \omega^4 - 2k^4\omega^2\eta^2 + k^8\eta^4 - 8k^2\omega^2\eta\varkappa|a_0|^2 < 0.$$

Let us take a look at the case $|a_0|^2 = 0$, that is a chain of infinite length. In this case $p$ takes the form:

$$p = \omega^2 + k^4\eta^2 \pm \sqrt{(\omega^2 - k^4\eta^2)^2} = \begin{cases} 2\omega^2 > 0, \\ 2k^4\eta^2 > 0. \end{cases}$$

Therefore in the case of an infinite chain all the roots of the equation (60) are real at any parameter values, that is in the case of $\omega' = 0$ a uniform distribution in an infinite chain is stable, as distinct from the case of $\omega' \neq 0$, when a uniform distribution in an infinite chain is unstable.

If $|a_0|^2 \neq 0$ then $p < 0$ at $k^2 < -2\frac{k}{\eta}|a_0|^2$, that is instability arises if $k/\eta < 0$ and $k < \sqrt{|2k|a_0|^2/\eta|}$ at any values of other parameters. But if $k/\eta > 0$, instability can arise at $q = \omega^4 - 2k^4\omega^2\eta^2 + k^8\eta^4 - 8k^2\omega^2\eta\varkappa|a_0|^2 < 0$. We have considered whether instability can arise in a finite ($|a_0|^2 \neq 0$) chain with DNA parameters: $\omega = 0.01, \eta = 1.276, \varkappa = 4$, having investigated numerically the expression $q(k,|a_0|^2)$ for these DNA parameters. Namely we have inquired into whether there exist k, at which the expression $q(k,|a_0|^2)$ has negative values. Our calculations suggest that the lowest negative value of $q(k,|a_0|^2)$ (for DNA parameters) vanishes, or, more precisely, goes up to positive values at $|a_0|^2 \to 0$, since the value of $q(k,|a_0|^2)$ at $|a_0|^2 = 0$ is positive for any values of k. It may be concluded that for a finite ($|a_0|^2 \neq 0$) chain there exist k, at which $q(k,|a_0|^2)$ has negative values. For example, the lowest negative value of $q(k,|a_0|^2) \approx -4 \cdot 10^{-11}$ for the chain length $N = 10^6$. Therefore, at $\omega' = 0$ in a finite chain with DNA parameters instability can arise. It can be shown that this instability leads to the formation of soliton state. Notice, that as distinct from [41], formation of a localized electron state occurs in a nonthreshold (in terms of the wave function's amplitude) manner.

The shorter is the chain, the greater is the absolute value of the lowest negative value of $q(k,|a_0|^2)$, both in the case of DNA parameters and in the case of any other values of the parameters $\omega, \eta$ and $\varkappa$. This means that the time in which instability arises becomes less. Under changes of the parameters $\omega, \eta$ and $\varkappa$, the general picture of the behavior of $q(k,|a_0|^2)$ does not change qualitatively.

## REFERENCES

1. Dekker C., Ratner M.A. *Phys.World.* 2001. V. 14. P. 29-34.

2. Boon E.M., Barton J.K. *Curr. Op. in Struct. Biol.* 2002. V. 12. P. 320-329

3. *Topics in Current Chemistry*. V. 236/237. Ed: Schuster G.B. Berlin:Springer-Verlag. 2004.

4. Keren K., Berman R.S., Buchstab E., Sivan U. and Braun E. *Science.* 2003. V. 302. P. 1380-1382

5. Lakhno V.D. *International Journal of Quantum Chemistry.* 2008. V. 108 P. 1970-1981

6. Lakhno V.D. *J.Biol. Phys.* 2000 V. 26. P. 133-147

7. Lakhno V.D., Fialko N.S. *R &C Dynamics.* 2002. V. 7 P. 299-313

8. Conwell E.M., Rakhmanova S.V. *Proc. Natl. Sci. U.S.A.* 2000 V. 97 P. 4556-4560

9. Lakhno V.D., Fialko N.S. *Phys. Lett. A.* 2000. V. 278 P. 108-111

10. Alexandre S.S., Artacho E., Soler J.M. and Chacham H. *Phys. Rev. Lett.* 2003. V. 91. 108105 (4 pages)

11. *Physics in one dimension*. Ed: Bernasconi J., Schneider T. Berlin:Springer-Verlag. 1981

12. Davydov A.S. *Sov. Phys. Usp.* 1982. V. 138 P. 603-643

13. Heeger A.J., Kivelson S. and Schrieffer J.R. *Rev. Mod. Phys.* 1988. V. 60 P. 781-850

14. Scott A. *Phys. Rep.* 1992. V. 217 P. 1-67

15. Okahata Y., Kobayashi T., Tanaka K. and Shimomura M.J. *Am. Chem. Soc.* 1992. V. 120 P. 6165-6166

16. Korshunova A.N., Lakhno V.D. *Mathematical Modeling* 2007. V. 19. P. 3-13

17. Lakhno V.D., Korshunova A.N. *Nelineynaya Dinamika*. 2008. V. 4. P. 193-214

18. Jian-Xin Zhu, Rasmussen K.O., Balatsky A.V. and Bishop A.R. *J. Phys.: Condens. Matter.* 2008. V. 19. P. 136203-136211

19. Kevrekidis P.G., Rasmussen K.O. and Bisop A.R. *Int. J. Modern Phys. B*. 2001. V. 15. P. 2833-2900

20. *Physica D. Special volume: Localization in Nonlinear Lattices* V. 119. Ed: Flach S., Mackay R.S. Amsterdam:Elsevier. 1998.

21. *Modern Methods for Theoretical Physical Chemistry of Biopolymers* Ed: Starikov E.B., Lewis J.P. and Tanaka S. Amsterdam:Elsevier. 2006.

22. Berashevich J.A., Bookatz A.D. and Chakraborty T. *J. Phys.: Condens. Matter.* 2008. V. 20. 035207 (5 pages)

23. Peyrard M. *Nonlinearity.* 2004. V. 17. P. R1-R40

24. Scott A. *Phys.Rep.* 1992. V. 217. P. 1-67

25. Kivshar Y.S., Agrawal G.P. In: *Optical Solitons. From Fibers to Photonic Crystals*. New York:Academic Press. 2003. 540 p.

26. Akhmediev N.N., Ankiewicz A.*Solitons: nonlinear pulses and beams* London:Chapman and Hall. 1997. 335 p.

27. Bountis (ed) 1991 *Collective Proton Transport in Hydrogen Bounded System, Proc. of the NATO Advanced Research Workshop (Heraclion, Crete)*, (New York, Plenum)

28. *Dissipative Solitons*. Lecture Notes in Physics, Vol. 661. Ed: Akhmediev N., Ankiewicz A. Berlin:Spinger. 2005. 448 p.

29. Holstein T. *Ann., Phys.* 1959. V. 8. P. 343-389

30. Lakhno V.D. and Fialko N.S. *Eur. Phys. J. B*. 2005. V. 43. P. 279-281

31. Carr L.D., Clark C.W. and Reinhardt W.P. *Phys. Rev. A.* 2000. V. 62. 063610 (10 pages)

32. Carr L.D., Clark C.W. and Reinhardt W.P. *Phys. Rev. A.* 2000. V. 62. 063611 (10 pages)

33. Vansant P., Smondyrev M.A., Peeters F.M. and Devreese J.T. *Zeitschrift Fur Physik B.* 1996. V. 99. P. 345-351

34. Starikov E.B. *Phil. Mag.* 2005. V. 85. P. 3435-3462

35. Porath D., Bezryadin A., de Vries S. and Dekker C. *Nature* 2000. V. 403. P. 635-638

36. Davydov A.S., Kislukha N.I. *Sov. Phys. JETP.* 1976. V. 44. P. 571-575

37. Davydov A.S. *Sov. Phys. Usp.* 1982. V. 25. P. 898-918

38. Hennig D., Chetverikov A.P., Velarde M.G. and Ebeling W. *Phys. Rev. E.* 2007. V. 76. 046602 (9 pages)

39. Lakhno V.D. In:*Modern Methods for Theoretical Physical Chemistry of Biopolymers* Ed: Starikov E.B., Lewis J.P. and Tanaka S. Amsterdam:Elsevier. 2006. P. 461-481.

40. Bykov V.P. *Uspekhi Fizicheskikh Nauk.* 2006. V. 176 P. 1007-1014

41. Kivshar Y.S. and Peyerard M. *Phys. Rev. A*. 1992. V. 46 P. 3198-3205